\documentclass[runningheads,a4paper]{llncs}
\usepackage{amssymb}
\usepackage{graphicx}
\usepackage{wrapfig}

\usepackage{url}
\urldef{\mailsa}\path|wenm@seas.upenn.edu|
\urldef{\mailsb}\path|utopcu@utexas.edu|

\usepackage{amsmath}

\usepackage{color}
\usepackage{algpseudocode}
\usepackage{algorithm}

\usepackage{cite}
\usepackage{kantlipsum}

\usepackage{pgf}
\usepackage{tikz}
\usetikzlibrary{arrows,automata}

\algnewcommand\True{\textbf{TRUE}\space}
\algnewcommand\False{\textbf{FALSE}\space}

\begin{document}

\mainmatter  

\title{Strategy Synthesis for Stochastic Rabin Games with Discounted Reward}

\titlerunning{Strategy Synthesis for Stochastic Rabin Games with Discounted Reward}

%
%
\author{Min Wen\inst{1}%
\and Ufuk Topcu\inst{2}}
%

\institute{$^{1}$Department of Electrical and Systems Engineering, University of Pennsylvania \\
\mailsa\\
$^{2}$Department of Aerospace Engineering and Engineering Mechanics, \\ University of Texas at Austin \\
\mailsb\\
}

%
%

\maketitle

\begin{abstract}
%
%

Stochastic games are often used to model reactive processes. We consider the problem of synthesizing an optimal almost-sure winning strategy in a two-player (namely a system and its environment) turn-based stochastic game with both a qualitative objective as a Rabin winning condition, and a quantitative objective as a discounted reward. Optimality is considered only over the almost-sure winning strategies, i.e., system strategies that guarantee the satisfaction of the Rabin condition with probability 1 regardless of the environment's strategy. We show that optimal almost-sure winning strategies may need infinite memory, but $\varepsilon$-optimal almost-sure winning strategies can always be finite-memory or even memoryless. We identify a sufficient and necessary condition of the existence of memoryless $\varepsilon$-optimal almost-sure winning strategies and propose an algorithm to compute one when this condition is satisfied. 

\end{abstract}

\section{Introduction}



Stochastic games, or $2\frac{1}{2}$-player graph games \cite{chatterjee2012survey}, are finite turn-based two-player games between a controlled \emph{system} and its uncontrolled \emph{environment} with probabilistic transitions. The state space is partitioned into system states and environment states, and each player can only take actions at its own states. Stochastic games are commonly used as models of reactive processes, where transition distributions encode the uncertainties in real executions. In reactive synthesis problems, $\omega$-regular languages are often considered as qualitative descriptions of the desired behaviors \cite{kress2007s}, which can be represented by, for example, Rabin objectives \cite{thomas1997languages}. Rabin objectives are described by a set of Rabin pairs in which each pair contains two disjoint subsets of states. An infinite path of the game is winning for the system if and only if there exists a Rabin pair such that all states in the first subset are visited only for finitely many times, and some states in the second subset are visited infinitely often. Despite this qualitative criterion, game paths can also be evaluated quantitatively with different reward functions. One classical and elegant reward function is the discounted reward \cite{shapley1953stochastic, littman1994markov, filar1996competitive}, which puts exponentially decaying weights to the rewards gained at different steps and therefore rewards gained in near future are weighed more than those gained in far future. The system aims to satisfy the Rabin objective or maximize the reward, while the environment is assumed to be adversarial and tries to violate the Rabin condition or minimize the reward.

In recent years there is an increasing interest in combining qualitative and quantitative objectives in reactive synthesis problems, as the two types of objectives serve different control purposes \cite{bloem2009better, chen2013synthesis, chatterjee2005mean, chatterjee2014perfect}. Intuitively, qualitative objectives like Rabin objectives act as task rules or functionality descriptions of the control system, while quantitative objectives like discounted rewards give a measure of how well the task is implemented. Combining the two types of objectives allows looking for near-optimal strategies for a given task.

We consider the strategy synthesis problems in stochastic games with both a Rabin objective and a discounted reward. Given a game and the objectives, we would like to synthesize a strategy for the system that is optimal or $\varepsilon$-optimal with respect to the discounted reward and guarantees the satisfaction of the Rabin objective with probability 1. Although the discounted reward majorly cares about finite-time performance, we consider it as a performance criterion of system strategies in the long run. As Rabin objectives are evaluated in infinite sequences, it makes more sense to evaluate system strategies at \emph{all} system states, especially those that are visited infinitely often, rather than only at the given initial state. The difference between these two cases can be illustrated by the example in Fig.~\ref{fig:discount}, for which we want to compute an $\varepsilon$-optimal strategy for the system. 
\begin{figure}[t!]
\minipage{0.52\textwidth}
\centering
\begin{tikzpicture}[->,>=stealth',shorten >=1pt,auto,node distance=2.5cm,
                    semithick]
  \tikzstyle{every state}=[draw=black,text=black]

  \node[initial,state,rectangle] (q0)                  {$q_0$};
  \node[state]         			(q1) [above of=q0]    {$q_1$};

  \path (q0)    edge [loop right]   node                {$(a_0, 0.999, 0)$} (q0)
                edge [bend left]    node[shift={(0,0)}] {$(a_0, 0.001, 0)$} (q1)
        (q1)    edge [loop right]   node[shift={(0,0)}] {$(a_1, 1, 1)$} (q1)
        		    edge [bend left]    node[shift={(0,0)}] {$(a_2, 1, 0)$} (q0);
\end{tikzpicture}
\caption{A stochastic game with Rabin pairs $\{(\emptyset, \{q_0\})\}$. $q_0$ is an environment state and $q_1$ is a system state. When the environment takes $a_0$ at $q_0$, the state transits to $q_1$ with probability 0.001 and does a self loop with probability 0.999. The reward is 0 in both cases. If the system takes $a_1$ at $q_1$, it does a self loop with probability 1 and the reward is 1; if it takes $a_2$ at $q_1$, the state transits back to $q_0$ with probability 1 and reward 0. }
\label{fig:discount}
\endminipage\hfill
\minipage{0.44\textwidth}
\centering
\begin{tikzpicture}[->,>=stealth',shorten >=1pt,auto,node distance=2.5cm,
                    semithick]
  \tikzstyle{every state}=[draw=black,text=black]

  \node[initial,state] (q0)                        {$q_0$};
  \node[state]         (q1) [above of=q0]    {$q_1$};
  \node[state]         (q2) [right of=q0]    {$q_2$};

  \path (q0)    edge                node                {$(a_0, 1, 0)$} (q1)
                edge                node[shift={(0,0)}] {$(a_1, 1, 0)$} (q2)
        (q1)    edge [loop right]   node[shift={(0,0)}] {$(a_2, 1, 1)$} (q1)
        (q2)    edge [loop above]   node[shift={(0,0)}] {$(a_3, 1, 0)$} (q2);
\end{tikzpicture}
\caption{A Rabin game $G$ with Rabin pairs $\{(\emptyset, \{q_2\})\}$ in which the optimal value in $\Sigma_{as}^G$ is strictly less than the optimal value in $\Sigma_s$. All transitions are deterministic and the only transition with positive reward is the self-loop by taking $a_2$ at $q_1$. The optimal value over $\Sigma_{as}^G$ at $s$ is 0, while the optimal value over $\Sigma_s$ at $s$ is $\frac{\gamma}{1-\gamma}$, where $\gamma$ is the discount factor.}
\label{fig:almostsureoptimal}
\endminipage\hfill
\end{figure}
Let $q_0$ be the initial state and the discount factor be $0.9$. With probability no less than 0.9, it takes the environment more than 100 steps to leave $q_0$ for the first time. This suggests that if we consider only the expected future discounted reward gained from the very first step at the initial state, the discounted reward is not adding any further constraints to the synthesis problem if $\varepsilon > \frac{0.9^{100}}{1-0.9} \approx 2.66e^{-4}$. However, if we consider the expected future discounted reward at all states, the discounted reward effectively encourages the system to stay in $q_1$ for at least $\frac{\log 0.1 \varepsilon}{\log 0.9}$ steps before it chooses to go back to $q_0$ to ensure the satisfaction of the Rabin objective. This second system strategy is more desirable as the system can recurrently gain high rewards while satisfying the qualitative requirements. 

Although in general, finite-memory $\varepsilon$-optimal almost-sure winning system strategies always exist and can be synthesized, we are particularly interested in the analysis and synthesis of memoryless solutions for the following three reasons. First, memoryless strategies suffice for the system for being both optimal and almost-sure winning (Theorem~\ref{thm:optimalsufficiency}). Second, memoryless $\varepsilon$-optimal strategies guarantee $\varepsilon$-optimal expected future discounted reward not only from any state and but also from any time in infinite game paths, while finite-memory $\varepsilon$-optimal strategies may only guarantee $\varepsilon$-optimality with initial memory at all states. Third, the reduced usage of memory keeps the solution simple and efficient.

\paragraph{Related work.} 
It is well-known that linear temporal logic (LTL) specifications can be transformed to deterministic Rabin automata \cite{babiak2013effective}, and there are existing tools to automate this procedure \cite{klein2005linear, gaiser2012rabinizer,  blahoudek2015ltl3dra, klein2015ltl2dstar}. Studies considering both the satisfaction of LTL specifications and optimization with respect to a quantitative objective use different models such as Markov decision processes \cite{wolff2012optimal, ding2014optimal}, non-deterministic systems \cite{wolff2013optimal} and nonlinear systems \cite{wolff2013nonlinear}. Less work has been dedicated to stochastic games. One piece of such work is by Chen et al. \cite{chen2013synthesis}, in which LTL specifications and the expected total reward, rather than the discounted reward, are considered. 

The work in \cite{de2003discounting, almagor2014discounting} connects LTL specifications with discounting and creates the so-called \emph{discounting LTL}, where the satisfaction of LTL specifications is evaluated quantitatively with discounting operators. Studies on discounting LTL focus on the model checking rather than synthesis. The initiatives of combining discounted rewards to encourage close future benefits are the same, but in our case, the discounted reward functions can be used to encode independent preferences from the Rabin objective, which makes the formulation more flexible. 

Most of the previous work on games with both qualitative and quantitative objectives is on mean-payoff parity games \cite{chatterjee2005mean, bloem2009better, chatterjee2014perfect}, which focus on the analysis of game values, (sure or almost-sure) winning regions and the class of strategies that suffices for the given objectives, rather than the synthesis of a system strategy, which is our major interest. 

\paragraph{Contribution.}
We analyze the synthesis of system strategies that both satisfy the given Rabin objective with probability 1, i.e. \emph{almost-sure winning}, and are optimal or $\varepsilon$-optimal with respect to the given discounted reward function. The discounted reward function is used as a performance criterion of system strategies, and the expected discounted reward is checked not only at initial states but also all other system states. Our main contributions are as follows. 

\begin{enumerate}
\item We show that memoryless strategies suffice for being both optimal and almost-sure winning for the system (Theorem~\ref{thm:optimalsufficiency}) and propose an algorithm to synthesize such a memoryless strategy if it exists (Algorithm~\ref{alg:p1}).
\item We show a sufficient and necessary condition of the existence of memoryless almost-sure winning system strategies that can be arbitrarily near-optimal (Theorem~\ref{thm:main}). Randomized strategies with distribution restrictions are used to reduce memory usage. 
\item If the previous condition is satisfied, we propose an algorithm (Algorithm~\ref{alg:suboptimal}) to synthesize an $\varepsilon$-optimal memoryless system strategy with any $\epsilon > 0$, which utilizes off-the-shelf algorithms for the synthesis of stochastic Rabin games. 
\end{enumerate}

\section{Preliminaries}
In this section we introduce the definitions and notations used in this paper. For any countable set $A$, denote its cardinality by $|A|$; denote be the sets of finite and infinite sequences composed of elements in $A$ by $A^{*}$ and $A^{\omega}$ respectively. Let $\mathcal{D}(A)$ be the set of all probability distributions defined on $A$. 

\paragraph{Turn-based Rabin game.}
A \emph{turn-based Rabin game} between the system and the environment is defined as a tuple $G = (S, S_s, S_e, I, A, T, W)$, where $S$ is a finite state space; $S_s \subseteq S$ is the set of states at which the system chooses actions, and $S_e := S \backslash S_s$ is the set of states at which the environment chooses actions; $I \subseteq S$ is the set of initial states; $A$ is a finite set of available actions; $T: S \times A \rightarrow \mathcal{D}(S)$ is the transition function; $W = \{ (E_1, F_1), \cdots, (E_d, F_d) \}$ is the set of Rabin pairs; $E_i, F_i \subseteq S$ and $E_i \bigcap F_i = \emptyset$ hold for all $i \in \{1, \cdots, d\}$.

Let $A^G: S \rightarrow 2^A \backslash \emptyset$ be a mapping from each state to its available actions in $G$. A turn-based Rabin game is \emph{deterministic} if for all $s \in S$ and $a \in A^G(s)$, $|\{s' \in S \mid T(s,a)(s') > 0\}| = 1$; otherwise it is \emph{probabilistic}. For a nonempty subset $S' \subseteq S$, we define the induction of a subgame from a subset as follows. 

\paragraph{Induced game.}
A nonempty set $S' \subseteq S$ \emph{induces a subgame $G'$} if it holds for all $s \in  S_e \bigcap S'$ and $a \in A^G(s)$ that $\{s' \mid T(s,a)(s') > 0\} \subseteq S'$, and for all $s \in S_s \bigcap S'$, there exists $a \in A^G(s)$ such that $\{s' \mid T(s,a)(s') > 0\} \subseteq S'$. We denote the induced subgame as $G' = G \upharpoonright S' = (S', S'_s, S'_e, I', A, T', W')$, where (1) $S'_s = S_s \bigcap S'$, $S'_e = S_e \bigcap S'$, $I' = I \bigcap S'$, $W' = \{ (E'_1, F'_1), \cdots, (E'_d, F'_d)\}$ such that $E'_i = E_i \bigcap S'$ and $F'_i = F_i \bigcap S'$ hold for all $i = 1, \cdots, d$; (2) for all $s \in S'$, $A^{G'}(s) = \{a \in A^G(s) \mid \{s' \in S \mid T(s,a)(s') > 0\} \subseteq S'\}$, i.e. by taking actions in $A^{G'}(s)$, the probability of entering $S \backslash S'$ is always zero; (3) for all $s \in S'$, $a \in A^{G'}(s)$, $T'(s,a)(s') = T(s,a)(s')$. 

A \emph{run} $\pi =  (s_{\pi}^0, a_{\pi}^1), (s_{\pi}^1, a_{\pi}^2), (s_{\pi}^2, a_{\pi}^3) \cdots := (s_{\pi}^{i-1}, a_{\pi}^i)_{i \in \mathbb{N}^+}$ of $G$ is an infinite sequence of state-action pairs such that for all $i \in \mathbb{N}^+$, $s_{\pi}^{i-1} \in S$ and $a_{\pi}^i \in A^G(s_{\pi}^i)$,  $T(s_{\pi}^{i-1} , a_{\pi}^i)(s_{\pi}^i) > 0$. Without loss of generality, assume that all states are \emph{reachable} from $I$ in $G$, i.e. for any state $s \in S$, there exists a run $\pi =  (s_{\pi}^{i-1}, a_{\pi}^i)_{i \in \mathbb{N}^+}$ and $k \in N$ such that $s_{\pi}^0 \in I$ and $s_{\pi}^k = s$. 

\paragraph{Strategy.}
A \emph{(randomized) strategy for the system} is defined as a tuple $\sigma_s = (\sigma^m_s, \rho^m_s, M_s, m^0_s)$, where $M_s$ is a (possibly countably infinite) set of memory states; $m^0_s \in M_s$ is the initial memory state; $\sigma^m_s: S_s \times M_s \rightarrow \mathcal{D}(A)$, and $\rho^m_s: S \times M_s \rightarrow M_s$ is the memory update function. If $M_s$ is a singleton, $\sigma_s$ is a \emph{memoryless} strategy; if $M_s$ is a finite set, $\sigma_s$ is a \emph{finite-memory strategy}. With a slight abuse of notation, we use $\sigma_s$ to represent $\sigma^m_s$ when $\sigma_s$ is memoryless. 
Let $A_{\sigma_s}: S_s \times M_s \rightarrow 2^A \backslash \emptyset$ be a map from each system state $s \in S_s$ with memory $m \in M_s$ to the set of actions allowed by $\sigma_s$, i.e. $A_{\sigma_s}(s, m) = \{ a \mid \sigma^m_s(s,m)(a) > 0, a \in A^G(s) \}$. If $\sigma_s$ is memoryless, we use $A_{\sigma_s}(s)$ to represent $A_{\sigma_s}(s,m^0_s)$.
If $| A_{\sigma_s}(s,m) | = 1$ for all $s \in S_s$ and $m \in M_s$, $\sigma_s$ is a \emph{deterministic} strategy. When the exact transition distribution is not of interest, we can define \emph{non-deterministic strategies} with $\sigma^m_s: S_s \times M_s \rightarrow 2^{A} \backslash \emptyset$.
A strategy $\sigma_e = (\sigma^m_e, \rho^m_e, M_e, m^0_e)$ for the environment can be defined analogously. Let $\Sigma_s$ and $\Sigma_e$ be the sets of all system strategies and environment strategies respectively. 

A run $\pi = (s_{\pi}^{i-1}, a_{\pi}^i)_{i \in \mathbb{N}^+}$ is \emph{feasible} for a pair of strategies $(\sigma_s, \sigma_e)$, where $\sigma_s = (\sigma^m_s, \rho^m_s, M_s, m^0_s)$ is a strategy for the system and $\sigma_e = (\sigma^m_e, \rho^m_e, M_e, m^0_e)$ is a strategy for the environment, if there exist sequences $(m_{\pi,s}^i)_{i \in \mathbb{N}} \in M_s^{\omega}$ and $(m_{\pi,e}^i)_{i \in \mathbb{N}} \in M_e^{\omega}$ such that (1) $m_{\pi,s}^0 = m^0_s$, $m_{\pi,e}^0 = m^0_e$; (2) for all $i \in \mathbb{N}$, $m_{\pi,s}^{i+1} = \rho^m_s(s_{\pi}^i, m_{\pi,s}^i)$, $m_{\pi,e}^{i+1} = \rho^m_e(s_{\pi}^i, m_{\pi,e}^i)$; (3) for all $i \in \mathbb{N}$ such that $s_{\pi}^i \in S_s$, $T(s_{\pi}^i, a_{\pi}^i)(s_{\pi}^{i+1}) > 0$ and $\sigma^m_s(s_{\pi}^i, m_{\pi,s}^i)(a_{\pi}^i) > 0$; (4) for all $i \in \mathbb{N}$ such that $s_{\pi}^i \in S_e$, $T(s_{\pi}^i, a_{\pi}^i)(s_{\pi}^{i+1}) > 0$ and $\sigma^m_e(s_{\pi}^i, m_{\pi,e}^i)(a_{\pi}^i) > 0$. Given a pair of strategies $(\sigma_s, \sigma_e)$ in the game $G$ and a state $s_0 \in S$, we denote the set of feasible runs starting at $s_0$ by $U_G(s_0, \sigma_s, \sigma_e)$. For all $\pi \in U_G(s_0, \sigma_s, \sigma_e)$, the probability that the pair of strategies lead to $\pi$ is denoted by $Pr^{(s_0, \sigma_s, \sigma_e)}(\pi)$. 


\paragraph{Winning region and winning strategy.} For a run $\pi$, let $Inf(\pi) \subseteq S$ be the set of states that are visited for infinitely many times in $\pi$. We say that $\pi$ is \emph{winning for the system} in a turn-based Rabin game $G = (S, S_s, S_e, I, A, T, W)$ if and only if there exists $i \in \{1, \cdots, d\}$ such that $Inf(\pi) \bigcap F_i \not= \emptyset$ and $Inf(\pi) \bigcap E_i = \emptyset$. The set of winning runs for the system is denoted by $\Phi(W)$. 
Given a turn-based Rabin game $G = (S, S_s, S_e, I, A, T, W)$ where $W = \{ (E_1, F_1), \cdots, (E_d, F_d) \}$, a strategy $\sigma_s$ for the system is \emph{sure winning} for the system if $U_G(s, \sigma_s, \sigma_e) \subseteq \Phi(W)$ holds at $s \in I$ for all environment strategy $\sigma_e$; a strategy $\sigma_s$ for the system is \emph{almost-sure winning} for the system if the conditional probability $Pr(\pi \in \Phi(W) \mid \pi \in U_G(s, \sigma_s, \sigma_e)) = 1$ at all $s \in I$. Let $\Sigma_{sw}^G$ and $\Sigma_{as}^G$ be the set of all sure winning and almost-sure winning system strategies in $G$, respectively. Also, let $W_{s}^G$ be the set of all states from which there exists a sure winning strategy for the system, which is called the \emph{sure winning region} of the system. The \emph{almost-sure winning region} $W_{as}^G$ of the system can be defined analogously.

\paragraph{Discounted reward.} Given a turn-based Rabin game $G = (S, S_s, S_e, I, A, T, W)$, an \emph{instantaneous reward function} is a mapping from transitions to their corresponding rewards $\mathcal{R}: S \times A \times S \rightarrow \mathbb{R}^{\geq 0}$. 
The \emph{discounted reward} for the system in a run $\pi = (s_{\pi}^{i-1}, a_{\pi}^i)_{i \in \mathbb{N}^+}$ of $G$, denoted by $\mathcal{J}_G(\pi)$, is a discounted sum of the instantaneous rewards it gains at each step, i.e., $\mathcal{J}_G({\pi}) = \sum_{t=0}^{\infty} \gamma^t \mathcal{R}(s_{\pi}^t, a_{\pi}^{t+1}, s_{\pi}^{t+1})$, where $\gamma \in (0,1)$ is a \emph{discount factor}. 

A \emph{value function} is a map $V_G: S \times \Sigma_s \times \Sigma_e \rightarrow \mathbb{R}^{\geq 0}$, which is the expected discounted reward gained by the system if the system takes the strategy $\sigma_s$ and the environment takes the strategy $\sigma_e$ from a state $s \in S$ in $G$, i.e., $V_G(s, \sigma_s, \sigma_e) = \sum_{\pi \in U_G(s_0, \sigma_s, \sigma_e)} Pr^{(s_0, \sigma_s, \sigma_e)}(\pi) \mathcal{J}_{G}(\pi)$. In particular, we define the \emph{value of a system strategy $\sigma_s$} for the system at $s \in S$ as $\tilde{V}_G(s, \sigma_s) = \inf_{\sigma'_e \in \Sigma_e} V_G(s, \sigma_s, \sigma'_e)$, which is the worst-case expected discounted reward the system can guarantee by taking the strategy $\sigma_s$. Therefore, the game is zero-sum for the system. 


\paragraph{Optimal strategy and $\varepsilon$-optimal strategy.} Given a set $\Sigma'_s \subseteq \Sigma_s$ of system strategies, a system strategy $\sigma_s \in \Sigma'_s$ is \emph{optimal over $\Sigma'_s$} if $\tilde{V}_G(s, \sigma_s) \geq \sup_{\sigma'_s \in \Sigma'_s} \tilde{V}_G(s, \sigma'_s)$ holds for all $s \in S$. The \emph{optimal value over $\Sigma'_s$} is a mapping $V^*_{\Sigma'_s}: S \rightarrow \mathbb{R}^{\geq 0}$ that maps each state $s$ to the value of optimal system strategies over $\Sigma'_s$, i.e., for all $s \in S$, $V^*_{\Sigma'_s}(s) = \sup_{\sigma'_s \in \Sigma'_s} \tilde{V}_G(s, \sigma'_s) = \sup_{\sigma'_s \in \Sigma'_s} \inf_{\sigma'_e \in \Sigma_e} V_G(s, \sigma'_s, \sigma'_e) = \inf_{\sigma'_e \in \Sigma_e} \sup_{\sigma'_s \in \Sigma'_s} V_G(s, \sigma'_s, \sigma'_e)$. For any $\varepsilon > 0$, a system strategy $\sigma_s$ is \emph{$\varepsilon$-optimal over $\Sigma'_s$} if $\tilde{V}_G(s, \sigma_s) \geq V^*_{\Sigma'_s}(s) - \varepsilon$ holds for all $s \in S$. If $\sigma_s$ is optimal ($\varepsilon$-optimal) over $\Sigma_s$, it is further called an \emph{optimal ($\varepsilon$-optimal) system strategy in $G$} and its value is called the \emph{optimal ($\varepsilon$-optimal) value for the system in $G$}. 
Optimal and $\varepsilon$-optimal strategies for the environment can be defined analogously. 


\paragraph{Sufficiency of a strategy class for an objective.} 
A class $\mathcal{C}$ of strategies for system \emph{suffices for an objective $\mathcal{O}$} if whenever there exists a strategy for the system satisfying the objective $\mathcal{O}$, there exists a strategy within the class $\mathcal{C}$ that also satisfies $\mathcal{O}$. The strategy classes discussed in this paper are deterministic strategies, non-deterministic strategies, randomized strategies, memoryless strategies, finite-memory strategies and their intersections. Objectives considered most in this paper are optimality, $\varepsilon$-optimality, and almost-sure winning for the system. 




\section{Problem Formulation}
\label{section:problemFormulation}
With the definitions and notations introduced in the previous section, we can now formulate the problems. We focus on optimal almost-sure winning strategies in the first problem. 
 

\begin{problem}
Given a zero-sum turn-based Rabin game $G^{in}$ in which the system has an almost-sure winning strategy, an instantaneous reward function $\mathcal{R}$ and a discount factor $\gamma \in (0,1)$, 
decide if there exists a finite-memory almost-sure winning strategy $\sigma^*_s$ that is optimal over all almost-sure winning strategies for the system. Synthesize one such $\sigma^*_s$ if it exists. 
\label{problem:existence}
\end{problem}

Note that Problem~\ref{problem:existence} considers the existence of an optimal system strategy over all almost-sure winning strategies in $\Sigma_{as}^G$ rather than over all system strategies in $\Sigma_s$, which is consistent with our motivation of considering discounted reward. The same sense of optimality is discussed in \cite{wen2015correct}. The optimal value over $\Sigma_{as}^G$ can be strictly less than that over $\Sigma_s$, which is illustrated in the example in Fig.~\ref{fig:almostsureoptimal}.

It has been shown in \cite{wen2015correct} that finite-memory strategies do not suffice for the objective of being both optimal and sure winning for the system in deterministic Rabin games with discounted rewards. As Problem~\ref{problem:existence} is even more general than the problem in \cite{wen2015correct}, finite-memory strategies do not suffice for the objectives in Problem~\ref{problem:existence}. If finite-memory strategies in Problem~\ref{problem:existence} do not exist in a specific problem, we consider an approximate solution instead, which is a finite-memory $\varepsilon$-optimal almost-sure winning strategy for the system with an arbitrary $\varepsilon > 0$. We formulate this problem as follows.

\begin{problem}
Given the inputs of Problem~\ref{problem:existence} and a constant $\varepsilon > 0$, synthesize a finite-memory almost-sure winning strategy $\sigma_{s,\varepsilon}$ for the system that is $\varepsilon$-optimal over all almost-sure winning system strategies, if $\sigma^*_s$ in Problem~\ref{problem:existence} does not exist. 
\label{problem:suboptimal}
\end{problem}


We solve the above two problems in the following steps. In Section~\ref{section:optimal} we show that memoryless strategies suffice for being both almost-sure winning and optimal over all almost-sure winning strategies for the system (Theorem~\ref{thm:optimalsufficiency}). We propose Algorithm~\ref{alg:p1} to solve Problem~\ref{alg:p1}. Then in section~\ref{section:nearoptimal} we show a sufficient and necessary condition of the existence of a memoryless almost-sure winning strategy for the system that is $\varepsilon$-optimal over all almost-sure winning strategies for all $\varepsilon > 0$ (Theorem~\ref{thm:main}). If this condition is satisfied, Algorithm~\ref{alg:suboptimal} can compute a (randomized) memoryless solution to Problem~\ref{problem:suboptimal} with any given $\varepsilon > 0$; otherwise Algorithm~\ref{alg:suboptimal} can get a finite-memory solution. 


\section{Optimal Almost-Sure Winning Strategies}
\label{section:optimal}

In this section, we concentrate on optimal almost-sure winning strategies for Problem~\ref{problem:existence}. 
As explained before, the optimality considered in Problem~\ref{problem:existence} is with respect to the values of all almost-sure winning strategies for the system. Intuitively, to solve Problem~\ref{problem:existence} we need to search for an optimal strategy over all strategies in $\Sigma_{as}^G$, but it is hard to encode all almost-sure winning system strategies compactly. However, Lemma~\ref{lemma:aswinningoptimal} and Lemma~\ref{lemma:optimalstrategy} allow searching for an almost-sure winning strategy over all optimal strategies of a newly constructed game. 
Based on this we prove the sufficiency of memoryless deterministic strategies for being both optimal and almost-sure winning, and propose an algorithm to solve Problem~\ref{problem:existence} if a solution exists. 



\paragraph{Optimal value over $\Sigma_{as}^G$ and over $\Sigma_s$.}
The example in Fig.~\ref{fig:almostsureoptimal} shows that the optimal value over the set $\Sigma_{as}^G$ of all almost-sure winning system strategies can be strictly less than that over the set $\Sigma_s$ of all system strategies. However, Lemma~\ref{lemma:aswinningoptimal} below shows that this can only happen when the almost-sure winning set $W_{as}^G$ is a proper subset of the state space $S$.

\begin{lemma}
Let $G = (S, S_s, S_e, I, A, T, W)$ be a zero-sum turn-based Rabin game, $\mathcal{R}$ be an instantaneous reward function and $\gamma \in (0,1)$ be a discount factor. 
If the almost-sure winning region $W_{as}^{G}$ for the system coincides with $S$, then the optimal value over $\Sigma_s$ is the same as the optimal value over $\Sigma_{as}^G$. 
\label{lemma:aswinningoptimal}
\end{lemma}


To prove Lemma~\ref{lemma:aswinningoptimal}, we borrow the results from \cite{chatterjee2005complexity, filar1996competitive} stated as Lemma \ref{lemma:PMsuffice}. 
By Lemma~\ref{lemma:PMsuffice}, the existence of almost-sure winning (respectively, optimal) strategies for the system guarantees the existence of a \emph{memoryless} almost-sure winning (respectively, optimal) strategy for the system. 

\begin{lemma}
\begin{enumerate}
\item \cite{chatterjee2005complexity} Deterministic memoryless strategies suffice for almost-sure winning with respect to Rabin objectives in turn-based stochastic games. 
\item \cite{filar1996competitive} Deterministic memoryless strategies suffice for optimality in zero-sum turn-based stochastic games with discounted rewards.
\end{enumerate}
\label{lemma:PMsuffice}
\end{lemma}

\begin{algorithm}[!t]
\begin{algorithmic}
\Function{FiniteMemStrategy}{$\sigma_s$, $\sigma'_s$, $C$}
    \State Build a new strategy $\sigma_{s,\varepsilon} = (\sigma^m_{s,\varepsilon}, \rho^m_{s,\varepsilon}, M_{s, \varepsilon}, m^0_{s,\varepsilon})$, where $M_{s, \varepsilon} = \{0, 1, \cdots, C\}$, $m^0_{s,\varepsilon} = 0$, 
    \[\rho^m_{s,\varepsilon}(s, m) = 
            \begin{cases}
                m+1,& \textbf{if } m < C, s \in S_s, \\
                m,& \textbf{otherwise.}
            \end{cases},
        \] and \[\sigma^m_{s,\varepsilon}(s,m)(a) = 
            \begin{cases}
                \sigma_s(s)(a),& \textbf{if }m \geq C, \\
                \sigma'_s(s)(a),& \textbf{otherwise}.
            \end{cases}
    \]
\label{alg:FiniteMemStrategy}
\EndFunction
\end{algorithmic}
\end{algorithm}

\paragraph{Proof sketch of Lemma~\ref{lemma:aswinningoptimal}.} 
As the state space $S$ of $G$ coincides with the almost-sure winning region $W_{as}^G$, Lemma~\ref{lemma:PMsuffice} guarantees the existence of a deterministic memoryless almost-sure winning strategy $\sigma_s$ for the system in $G$. 
Lemma~\ref{lemma:PMsuffice} also guarantees that the system always has a deterministic memoryless optimal system strategy $\sigma'_s$, as there always exist optimal system strategies with respect to Rabin objectives in turn-based games \cite{filar1996competitive}. 
Then for any nonnegative integer $C$, the system strategy $\sigma_s(C) := \textbf{FiniteMemStrategy}(\sigma_s, \sigma'_s, C)$ is almost-sure winning at all states in $W_{as}^G = S$. As $C$ increases, the value of $\sigma_s(C)$ approaches that of $\sigma'_s$, which is the optimal value over $\Sigma_s$. As the optimal value over $\Sigma_{as}^G$ is no less than the supremum of the values of $\sigma_s(C)$ for all $C$, we know the optimal value over $\Sigma_{as}^G$ coincides with that over $\Sigma_s$.


\paragraph{Sufficient and necessary condition of optimality.}
We then consider a sufficient and necessary condition for a system strategy to be optimal in zero-sum turn-based stochastic games with discounted rewards. 
Such games always have optimal value functions \cite{filar1996competitive}, which equal the unique solution of \eqref{eq:optimal_value_function} for all $s \in S$. 
\begin{equation}
V^*(s) = 
\begin{cases}
\max_{a \in A^G(s)} \sum_{s'\in S} T(s,a)(s') \big( \mathcal{R}(s,a,s') + \gamma V^*(s') \big) &\text{if }s \in S_s, \\
\min_{a \in A^G(s)} \sum_{s'\in S} T(s,a)(s') \big( \mathcal{R}(s,a,s') + \gamma V^*(s') \big) &\text{if }s \in S_e.
\end{cases}
\label{eq:optimal_value_function}
\end{equation}

Also, the set of optimal memoryless strategies for the system is exactly those $\sigma'_s$ satisfying the following conditions for all $s \in S_s$ \cite{filar1996competitive}: 
\begin{equation}
A_{\sigma'_s}(s) \subseteq \arg\max_{a \in A^G(s)} \sum_{s'\in S} T(s,a)(s') \big( \mathcal{R}(s,a,s') + \gamma V^*(s') \big). 
\label{eq:optimal_strategy}
\end{equation}
We call $A^*(s) := \arg\max_{a \in A^G(s)} \sum_{s'\in S} T(s,a)(s') \big( \mathcal{R}(s,a,s') + \gamma V^*(s') \big)$ the set of \emph{optimal actions} at $s \in S_s$, and actions in $A^G(s) \backslash A^*(s)$ \emph{suboptimal actions} at $s$. Therefore a memoryless system strategy $\sigma_s$ is optimal if and only if $A_{\sigma_s}(s) \subseteq A^*(s)$ holds for all $s \in S_s$. Furthermore, a finite-memory system strategy $\sigma_s = (\sigma_s^m, \rho_s^m, M_s, m_s^0)$ is optimal if and only if $A_{\sigma_s^m}(s, m) \subseteq A^*(s)$ holds for all $s \in S_s$ and $m \in M_s$. Assume $\sigma_s$ is a finite-memory optimal strategy, then for all $s \in S_s$ and $m \in M_s$, the expected discounted reward at $(s,m)$ must be the same as $V^*(s)$, the optimal value of $s$. Therefore the optimal actions at $(s,m)$ can only be a subset of $A^*(s)$. This key property of optimal strategies is summarized in Lemma~\ref{lemma:optimalstrategy}.

\begin{lemma}
Given a zero-sum turn-based Rabin game $G = (S, S_s, S_e, I, A, T, W)$, an instantaneous reward function $\mathcal{R}$ and a discount factor $\gamma \in (0,1)$, a system strategy $\sigma_s = (\sigma_s^m, \rho_s^m, M_s, m_s^0)$ is optimal with respect to the discounted reward if and only if $A_{\sigma_s}(s,m) \subseteq A^*(s)$ holds for all $s \in S_s$ and $m \in M_s$. 
\label{lemma:optimalstrategy}
\end{lemma}

As a result, if we limit the set of available actions at $s \in S_s$ to be $A^*(s)$, we can construct a new game $\bar{G}$ such that a system strategy $\sigma_s$ is optimal in $G$ if and only if it is \emph{a strategy} for the system in $\bar{G}$. 

\paragraph{Synthesis of optimal almost-sure winning strategies.} With the previous analysis we are now ready to propose an Algorithm~\ref{alg:p1} to solve Problem~\ref{problem:existence}. 


As we assume that the system has an almost-sure winning strategy in $G^{in}$, the almost-sure winning region $W_{as}^{G_{in}}$ is nonempty. Computation of almost-sure winning regions and strategies in stochastic Rabin games can be performed with off-the-shelf algorithms as those in \cite{chatterjee2005complexity, chatterjee2006strategy}, and we omit the details here. 
We can construct a subgame $G$ of $G^{in}$ such that $G = G^{in} \upharpoonright W_{as}^{G^{in}}$. By definition of almost-sure winning region and induced game, $\Sigma_{as}^{G^{in}} = \Sigma_{as}^G$, i.e. all almost-sure winning strategies in $G^{in}$ are preserved in $G$; by Lemma~\ref{lemma:aswinningoptimal}, the optimal value over $\Sigma_{as}^G$ is the same as that over $\Sigma_s$. 
Therefore a solution to Problem~\ref{problem:existence} must be both almost-sure winning and optimal over all system strategies in $G$. 

Then we utilize existing methods like value iteration to compute the optimal value function $V^*$ of $G$. With $V^*$ we compute the optimal actions $A^*(s)$ for all $s \in S_s$. Lemma~\ref{lemma:optimalstrategy} suggests that we can construct a new game $\bar{G}$ from $G$ by forcing the system to take only optimal actions, such that a system strategy is optimal in $G$ if and only if it is a system strategy in $\bar{G}$. In other words, exactly the set of all optimal system strategies are preserved in $\bar{G}$. 

Therefore, an almost-sure winning strategy $\sigma_s$ is optimal over all almost-sure winning system strategies in $G^{in}$ if and only if $\sigma_s$ is both almost-sure winning for the system and optimal in $G$, which is further equivalent to being almost-sure winning in $\bar{G}$. Hence any solution $\sigma_s$ to Problem~\ref{problem:existence} is an almost-sure winning strategy in $\bar{G}$, and vice versa. 

By Lemma~\ref{lemma:PMsuffice}, deterministic memoryless strategies suffice for almost-sure winning with Rabin objectives on turn-based stochastic games, and we end up with the following theorem. 

\begin{theorem}
Given a zero-sum turn-based Rabin game $G$, an instantaneous reward function $\mathcal{R}$ and a discount factor $\gamma \in (0,1)$, 
deterministic memoryless strategies suffice for being both almost-sure winning and optimal over all almost-sure winning strategies for the system. 
\label{thm:optimalsufficiency}
\end{theorem}

\begin{algorithm}[!t]
\begin{algorithmic}[1]
\Require {A turn-based Rabin game $G^{in} = (S^{in}, S^{in}_s, S^{in}_e, I^{in}, A^{in}, T^{in}, W^{in})$ in which the system has an almost-sure winning strategy, an instantaneous reward function $\mathcal{R}$, and a discount factor $\gamma \in (0, 1)$.}
\Ensure {\True if there exists a finite-memory almost-sure winning strategy $\sigma_s^*$ for the system that is optimal over all almost-sure winning strategies for the system; \False otherwise. }
    \State Compute the almost-sure winning region $W_{as}^{G^{in}}$ for the system in $G^{in}$. 
    \State Construct a new game $G = G^{in} \upharpoonright W_{as}^{G^{in}} = (S, S_s, S_e, I, A, T, W)$. 
    \State Compute the optimal value function $V^*$ for $G$. 
    \State Compute the optimal actions $A^*(s) := \arg\max_{a' \in A^G(s)} \Big( \sum_{s'} T(s, a')(s') \big( \mathcal{R}(s,a',s') + \gamma V^*(s') \big) \Big)$ for all $s \in S_s$.
	\State Construct a new game $\bar{G} = (\bar{S}, \bar{S}_s, \bar{S}_e, \bar{I}, A, T, \bar{W})$ such that $A^{\bar{G}}(s) = A^*(s)$ for all $s \in S_s$. $\bar{S}, \bar{S}_s, \bar{S}_e, \bar{I}$ and $\bar{W}$ are the reachable subsets of their corresponding component in $G$.
    \State Compute the almost-sure winning region ${W}_{as}^{\bar{G}}$ for the system in $\bar{G}$. 
    \If{$I \subseteq {W}_{as}^{\bar{G}}$} 
    		\State Compute a deterministic memoryless almost-sure winning strategy $\sigma_s$ for the system in $\bar{G}$. $\sigma^*_s \gets \sigma_s$.
        \State \Return \True and $\sigma^*_s$ is a solution to Problem~\ref{problem:existence}.
    \Else
    		\State \Return \False.
    \EndIf
\end{algorithmic}
\caption{Pseudo algorithm for Problem~\ref{problem:existence}.}
\label{alg:p1}
\end{algorithm}

\section{Near-Optimal Almost-Sure Winning Strategies}
\label{section:nearoptimal}

%
%
%


We showed in Section~\ref{section:problemFormulation} that finite-memory strategies do not suffice for the objectives in Problem~\ref{problem:existence}. For cases in which finite-memory solutions to Problem~\ref{problem:existence} do not exist, we relax the optimality objective and consider near-optimal almost-sure winning strategies for the system as stated in Problem~\ref{problem:suboptimal}, and Proposition~\ref{prop:eoptimalfinitesufficient} shows that finite-memory strategies for the system suffice for these objectives. 

\begin{proposition}
Given a zero-sum turn-based Rabin game $G$, an instantaneous reward function $\mathcal{R}$ with upper bound $R_{max}$ and a discount factor $\gamma \in (0,1)$, if $\sigma_s$ is a memoryless almost-sure winning strategy and $\sigma'_s$ is a memoryless optimal system strategy, then for any $\varepsilon > 0$ and $C > \frac{\log(\varepsilon (1-\gamma) / R_{max})}{\log(\gamma)}$, $\sigma_{s, \varepsilon}^f := \textbf{FiniteMemStrategy}(\sigma_s, \sigma'_s, C)$ is an $\varepsilon$-optimal almost-sure winning strategy for the system. 
\label{prop:eoptimalfinitesufficient}
\end{proposition}

The finite-memory strategy $\sigma_{s, \varepsilon}^f$ is a solution to Problem~\ref{problem:suboptimal}, but it may not be desirable compared with memoryless solutions (if exist), for two main reasons.
First, the number of memory states $|M_s|$ in $\sigma_{s, \varepsilon}^f$ grows linearly in $\log (1 \backslash \varepsilon)$. 
The execution of $\sigma_{s, \varepsilon}^f$ is equivalent to taking a memoryless strategy in a game in which the number of states is $|M_s|$ times the number of states in the original game $G$. 
Second, finite-memory $\varepsilon$-optimal strategies can only guarantee $\varepsilon$-optimal discounted reward at the start of runs, while memoryless $\varepsilon$-optimal strategies can guarantee $\varepsilon$-optimal future discounted reward from any step during the infinite execution. For all these reasons, we focus on \emph{memoryless} $\varepsilon$-optimal almost-sure winning strategies in this section. 






\paragraph{$\varepsilon$-optimal memoryless strategies.}
First we check the existence of $\varepsilon$-optimal memoryless system strategies for any $\varepsilon > 0$, without considering the almost-sure winning objective. 
We show that for all $\varepsilon > 0$, $\varepsilon$-optimal randomized memoryless system strategies always exist. 


Lemma~\ref{lemma:optimalstrategy} guarantees that a system strategy is optimal if and only if the actions allowed at all system states are always optimal actions. Allowing the system to take suboptimal actions will result in suboptimal strategies, but given the upper bound of the instantaneous reward, we can bound the suboptimality of the memoryless system strategy by restricting the probability that the system takes suboptimal actions at each system state. This key observation is summarized in the following lemma. 

\begin{lemma}
Let $G$ be a turn-based game, $\mathcal{R}$ be an instantaneous reward function with upper bound $R_{max}$, and $\gamma$ be the discount factor. For any $\varepsilon > 0$, if the probability that the system chooses an optimal action is at least $1 - \frac{(1-\gamma)^2 \varepsilon}{R_{max} - \varepsilon(1-\gamma)\gamma}$ at all system states, the strategy for the system is $\varepsilon$-optimal.
\label{lemma:eoptimal}
\end{lemma}

As optimal actions exist at all system states, Lemma~\ref{lemma:eoptimal} proves the existence of $\varepsilon$-optimal memoryless system strategies for all $\varepsilon > 0$. Therefore finite-memory strategies suffice for $\varepsilon$-optimality for all $\varepsilon > 0$. 


\paragraph{Independence of almost-sure winning on distributions.}

Now we consider memoryless $\varepsilon$-optimal system strategies that are almost-sure winning with respect to a Rabin objective. 
The following lemma shows that, whether a memoryless system strategy $\sigma_s$ is almost-sure winning or not is independent of the exact distribution $\sigma_s(s)$, if $A_{\sigma_s}(s)$ is given for all system state $s$. 


%
%
%

\begin{lemma}
Let a turn-based Rabin game $G = (S, S_s, S_e, I, A, T, W)$ be given. Let $\sigma_s^1$ and $\sigma_s^2$ be two memoryless strategies for the system in $G$. 
Provided that $A_{\sigma_s^1}(s) = A_{\sigma_s^2}(s)$ holds for all $s \in S_s$, $\sigma_s^1$ is almost-sure winning for the system if and only if $\sigma_s^2$ is almost-sure winning for the system.
\label{lemma:aswinning}
\end{lemma}

The proof idea of Lemma~\ref{lemma:aswinning} is similar to that of Theorem 3 in \cite{chatterjee2005complexity}. It has been shown that with probability $1$ the set of states that are visited infinitely often in an infinite run is an \emph{end component} \cite{de1997formal}, which is a strongly connected subset of $S$ from which there are no outgoing transitions for both players. Assume $W = \{(E_1, F_1), \cdots, (E_d, F_d)\}$, then a system strategy $\sigma_s$ is almost-sure winning if and only if, by taking $\sigma_s$, there exists $i \in \{1, \cdots, d\}$ for each reachable end component $U \subseteq S$ such that $U \bigcap E_i = \emptyset$ and $U \bigcap F_i \not= \emptyset$, regardless of the environment strategy. If $A_{\sigma_s^1}(s) = A_{\sigma_s^2}(s)$ holds for all $s \in S$, then, with any environment strategy $\sigma_e$, the sets of reachable end components are the same for the two strategy pairs $(\sigma_s^1, \sigma_e)$ and $(\sigma_s^2, \sigma_e)$. Therefore the two strategies $\sigma_s^1$ and $\sigma_s^2$ can only be almost-sure winning simultaneously. 

%
%

\paragraph{Connecting the two objectives.}
Lemma~\ref{lemma:eoptimal} and Lemma~\ref{lemma:aswinning} suggest that the two objectives in Problem~\ref{problem:suboptimal} can be partially decoupled in the synthesis of a memoryless system strategy $\sigma_s$: in order to be almost-sure winning, we only need to set $A_{\sigma_s}$ properly; and in order to be $\varepsilon$-optimal, we only need to make sure that $A_{\sigma_s}(s) \bigcap A^*(s) \not= \emptyset$ and $\sum_{a \in A_{\sigma_s}(s)} \sigma_s(s)(a)$ is bounded properly for all $s \in S_s$. If these conditions can be satisfied at the same time, $\sigma_s$ is memoryless, almost-sure winning and $\varepsilon$-optimal. We summarize this result in the following theorem. 

\begin{theorem}
Let a turn-based game $G = (S, S_s, S_e, I, A, T, W)$, an instantaneous reward function $\mathcal{R}$ and a discount factor $\gamma \in (0,1)$ be given. If there exists a memoryless almost-sure winning strategy $\sigma_s$ for the system that allows taking optimal actions at all system states, 
i.e., for all $s \in S_s$, $A_{\sigma_s}(s) \bigcap A^*(s) \not= \emptyset$, then the class of memoryless strategies for the system suffices for the objective of being both almost-sure winning and $\varepsilon$-optimal for all $\varepsilon > 0$.
\label{thm:arbitrarye}
\end{theorem}

It is possible that the conditions in Lemma~\ref{lemma:eoptimal} and Lemma~\ref{lemma:aswinning} cannot be satisfied at the same time. In such cases, there exists some positive $\varepsilon' > 0$ such that memoryless $\varepsilon'$-optimal almost-sure winning strategies for the system do not exist. 

\begin{lemma}
Let a turn-based game $G = (S, S_s, S_e, I, A, T, W)$, an instantaneous reward function $\mathcal{R}$ and a discount factor $\gamma \in (0,1)$ be given. If, for any memoryless almost-sure winning strategy $\sigma_s$ for the system, there exists a state $s \in S_s$ such that no optimal action can be taken, i.e., $A_{\sigma_s}(s) \bigcap A^*(s) = \emptyset$, then there exists $\varepsilon' > 0$ such that no memoryless strategies for the system can be both $\varepsilon'$-optimal and almost-sure winning.
\label{lemma:epsilonno}
\end{lemma}

\begin{figure}[t!]
\centering
\begin{tikzpicture}[->,>=stealth',shorten >=1pt,auto,node distance=2.8cm,
                    semithick]
  \tikzstyle{every state}=[draw=black,text=black]

  \node[initial,state] (q0)                  {$s_0$};
  \node[state]         (q1) [right of=q0]    {$s_1$};

  \path (q0)    edge [loop above]   node                {$(a_0, 0, 1)$} (q0)
                edge [bend left]    node[shift={(0,0)}] {$(a_1, 0, 1)$} (q1)
        (q1)    edge [loop above]   node[shift={(0,0)}] {$(a_2, 2, 1)$} (q1)
                edge [bend left]    node[shift={(0,0)}] {$(a_3, 0, 1)$} (q0);
\end{tikzpicture}
\caption{A turn-based Rabin game $G$ in which the system does not have memoryless almost-sure $\varepsilon$-optimal strategy for all $\varepsilon$. The labels on transitions show the underlying actions, the instantaneous rewards and the transition probabilities. The unique Rabin pair is $(\{s_1\}, \{s_0\})$. }
\label{fig:counterexample}
\end{figure}
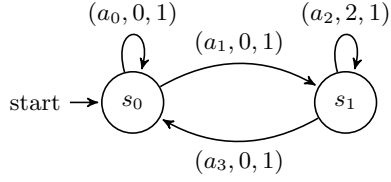

The condition in Lemma~\ref{lemma:epsilonno} can be illustrated by the example in Fig.~\ref{fig:counterexample}, in which the system can only win by visiting $s_0$ infinitely often and visiting $s_1$ for finitely many times. The optimal actions are $A^*(s_0) = \{a_1\}$, $A^*(s_1) = \{a_2\}$. 
Both states belong to the almost-sure winning region. If a memoryless strategy for the system is to guarantee almost-sure winning, it cannot allow taking $a_1$ at $s_0$, i.e., no optimal actions can be allowed at $s_0$. 
Then if $\varepsilon < \frac{2 \gamma}{1-\gamma}$,
the system does not have \emph{memoryless} $\varepsilon$-optimal almost-sure winning strategies for the system. 




Theorem~\ref{thm:arbitrarye} and Lemma~\ref{lemma:epsilonno} can be combined into the following main theorem, which is a sufficient and necessary condition of the existence of memoryless solutions to Problem~\ref{problem:suboptimal} for all $\varepsilon > 0$ simultaneously. 

\begin{theorem}
Given a turn-based game $G = (S, S_s, S_e, I, A, T, W)$, an instantaneous reward function $\mathcal{R}$ and a discount factor $\gamma \in (0,1)$, the class of memoryless strategies for the system suffices for the objective of being both almost-sure winning and $\varepsilon$-optimal for all $\varepsilon > 0$ simultaneously if and only if there exists a memoryless almost-sure winning system strategy $\sigma_s$ that allows taking optimal actions at all states in $S_s$, i.e., for all $s \in S_s$, $A_{\sigma_s}(s) \bigcap A^*(s) \not= \emptyset$.
\label{thm:main}
\end{theorem}

\paragraph{Synthesis of optimal almost-sure winning strategies.} 
We now propose Algorithm~\ref{alg:suboptimal} to solve Problem~\ref{problem:suboptimal}. If the condition in Theorem~\ref{thm:arbitrarye} is satisfied, the strategy synthesized by Algorithm~\ref{alg:suboptimal} is memoryless; otherwise it is finite-memory. In the second case, a finite-memory solution can be computed by the function {\bf FiniteMemStrategy}, as explained in Proposition~\ref{prop:eoptimalfinitesufficient}. Here we focus on the synthesis of a memoryless solution. 

\begin{algorithm}[!t]
\begin{algorithmic}[1]
\Require{A turn-based Rabin game $G^{in} = (S^{in}, S^{in}_s, S^{in}_e, I^{in}, A^{in}, T^{in}, W^{in})$ in which the system has an almost-sure winning strategy, an instantaneous reward function $\mathcal{R}$ with upper bound $R_{max}$, a discount factor $\gamma \in (0, 1)$ and a constant $\varepsilon > 0$.}
\Ensure {An $\varepsilon$-optimal almost-sure winning strategy $\sigma_{s,\varepsilon}$ for the system. }
    \State Compute the almost-sure winning region $W_{as}^{G^{in}}$ for the system in $G^{in}$. Let $\sigma'_s$ be a deterministic memoryless almost-sure winning strategy for the system in $G$.
    \State \label{step:G}Construct a subgame of $G^{in}$ as $G = G^{in} \upharpoonright W_{as}^{G^{in}} = (S, S_s, S_e, I, A, T, W)$. 
    \State Compute the optimal value function $V^*$ for $G$. 
    \State Compute the optimal actions $A^*(s) := \arg\max_{a' \in A^G(s)} \Big( \sum_{s'} T(s, a')(s') \big( \mathcal{R}(s,a',s') + \gamma V^*(s') \big) \Big)$ for all $s \in S_s$.
    \State \label{step:hatG}Construct a new game $\hat{G} = (\hat{S}, \hat{S}_s, \hat{S}_e, I, \hat{A}, \hat{T}, W)$, where $\hat{S} = \hat{S}_e \bigcup \hat{S}_s$, $\hat{S}_s = S_s \bigcup S_s^n \bigcup S_s^g$,  $\hat{S}_e = S_e \bigcup S_e^e$; $\hat{A} = A \bigcup \{\hat{a}\}$; $\hat{T}$ is defined in \eqref{eq:transitionFunction}. 
    \State \label{step:ashatG}Compute the almost-sure winning region ${W}_{as}^{\hat{G}}$ and a memoryless almost-sure winning strategy $\hat{\sigma}_s$ for the system in $\hat{G}$. 
    \If{$I \subseteq {W}_{as}^{\hat{G}}$} 
        \State \label{step:memoryless}Construct $\sigma_s$ as in \eqref{eq:construct1}, $\sigma_{s,\varepsilon} \gets \sigma_s$.
    \Else
        \State Compute a deterministic memoryless optimal system strategy $\bar{\sigma}'_s$ in $G'$.
        \State $C \gets \frac{\log(\varepsilon (1-\gamma) / R_{max})}{\log(\gamma)}$, $\sigma_{s, \varepsilon} \gets \textbf{FiniteMemStrategy}(\sigma'_s, \bar{\sigma}'_s, C)$. 
    \EndIf
    \State \Return $\sigma_{s,\varepsilon}$.
\end{algorithmic}
\caption{Pseudo algorithm for Problem~\ref{problem:suboptimal}.}
\label{alg:suboptimal}
\end{algorithm}

Let $G$ be the game constructed in Step~\ref{step:G} of Algorithm~\ref{alg:suboptimal}, and the optimal value function be $V^*$. As the state space of $G$ coincides with the $W_{as}^{G}$, Lemma~\ref{lemma:aswinningoptimal} guarantees that $V^*$ is also the optimal value function over all almost-sure winning strategies in $G^{in}$. Therefore an $\varepsilon$-optimal almost-sure winning system strategy in $G$ is also $\varepsilon$-optimal and almost-sure winning in $G^{in}$, and vice versa. As a result, we can synthesize a solution to Problem~\ref{problem:suboptimal} in $G$ instead of $G^{in}$.

Assume that the condition in Theorem~\ref{thm:arbitrarye} is satisfied. In order to compute a memoryless $\varepsilon$-optimal almost-sure winning strategy $\sigma_{s,\varepsilon}$, we need to consider both the constraint on $A_{\sigma_{s,\varepsilon}}(s)$ and the probability bound on $A^*(s)$ for all $s \in S_s$ at the same time. 
The approach in Algorithm~\ref{alg:suboptimal} is to construct a new turn-based Rabin game $\hat{G}$ from $G$ (Step~\ref{step:hatG}) that satisfies the following lemma. 


\begin{lemma}
Let $G$ and $\hat{G}$ be the turn-based Rabin games constructed in Step~\ref{step:G} and Step~\ref{step:hatG} of Algorithm~\ref{alg:suboptimal} respectively. $\varepsilon > 0$ is an input of Algorithm~\ref{alg:suboptimal}. 
Then each memoryless system strategy $\hat{\sigma}_s$ in $\hat{G}$ can be used to construct a memoryless $\varepsilon$-optimal system strategy $\sigma_s$ in $G$ such that
\begin{itemize}
\item $\sigma_s$ is $\varepsilon$-optimal in $G$; and
\item $\hat{\sigma}_s$ is almost-sure winning for the system in $\hat{G}$ if and only if the constructed $\sigma_s$ is almost-sure winning for the system in $G$. 
\end{itemize}
\label{lemma:prop1}
\end{lemma}



Given the two properties of $\hat{G}$ in Lemma~\ref{lemma:prop1}, 
it suffices to compute a memoryless almost-sure winning strategy for the system in $\hat{G}$ (Step~\ref{step:ashatG}) in order to compute a memoryless \emph{$\varepsilon$-optimal} almost-sure winning strategy for the system in $G$, which can be solved again with off-the-shelf algorithms \cite{chatterjee2005complexity, chatterjee2006strategy}. 


We now show the construction of $\hat{G} = (\hat{S}, \hat{S}_s, \hat{S_e}, I, \hat{A}, \hat{T}, W)$ from $G = (S, S_s, S_e, I, A, T, W)$ and verify that it satisfies Lemma~\ref{lemma:prop1}. 
Let $\hat{S}_s = S_s \bigcup S_s^n \bigcup S_s^g$, $\hat{S}_e = S_e$, and $\hat{S} = \hat{S}_e \bigcup \hat{S}_s$. $S_s^g$ and $S_s^n$ are two sets of new system states that are mutually disjoint. Let $O: S_s \rightarrow S_s^o$ and $N: S_s \rightarrow S_s^n$ be two bijective functions, and we use $O^{-1}$ and $N^{-1}$ to denote their inverse functions. For each state $s \in S_s$ in $G$, we add two states $O(s) \in S_s^o$ and $N(s) \in S_s^n$ to $\hat{S}_s$. 
The set of available actions at each state $s \in \hat{S}$ is defined as
\begin{equation}
\hat{A}(s) = 
\begin{cases}
\{\hat{a}\} &\text{ if }s \in S_s, \\
A^G(N^{-1}(s)) &\text{ if }s \in S_s^n, \\
A^*(O^{-1}(s)) &\text{ if }s \in S_s^o, \\
A^G(s) &\text{ if }s \in S_e.
\end{cases}
\label{eq:actionSet}
\end{equation}
The transition function $\hat{T}: \hat{S} \times \hat{A} \rightarrow \mathcal{D}(\hat{S})$ is defined  in \eqref{eq:transitionFunction}, where each transition from a state $s \in S_s$ in $G$ is separated into two transitions in $\hat{G}$. From each $s \in S_s$, there is only one available action $\hat{a}$, which transits from $s$ to $N(s)$ with probability $p := \frac{(1-\gamma)^2 \varepsilon}{R_{max} - \varepsilon(1-\gamma)\gamma}$ and to $O(s)$ with probability $1-p$. If it transits to $N(s)$, the system is free to choose from all actions in $A^G(s)$; otherwise the system can only take an optimal action in $A^*(s)$. The transition distribution $\hat{T}(s,a)$ for the second transition where $s' \in N(s) \bigcup O(s)$ is the same as $T(s, a)$ for all $a \in \hat{A}(s')$. 
\begin{equation}
\hat{T}(s,a)(s') = 
\begin{cases}
p &\text{ if }s \in S_s, s'=N(s),\\
1-p &\text{ if }s \in S_s, s'=O(s),\\
T(N^{-1}(s),a)(s') &\text{ if }s \in S_s^n, \\
T(O^{-1}(s),a)(s') &\text{ if }s \in S_s^o, \\
T(s,a)(s') &\text{ if }s \in S_e.
\end{cases}
\label{eq:transitionFunction}
\end{equation}

With each memoryless system strategy $\hat{\sigma}_s$ in $\hat{G}$, we can construct a memoryless system strategy $\sigma_s$ in $G$ such that for all $s \in S_s$ and $a \in A^G(s)$, 
\begin{equation}
\sigma_s(s)(a) = 
\begin{cases}
p \hat{\sigma}_s(N(s))(a) + (1-p)\hat{\sigma}_s(O(s))(a) &\text{ if }a \in A^*(s), \\
p \hat{\sigma}_s(N(s))(a) &\text{ if }a \not\in A^*(s). 
\end{cases}
\label{eq:construct1}
\end{equation}
This two-step decomposition of system transitions in $\hat{G}$ ensures that when the system takes $\sigma_s$, the probability of taking suboptimal actions at each system state is bounded by $p$.  
By Lemma~\ref{lemma:eoptimal}, $\sigma_s$ is $\varepsilon$-optimal. The fact that $\sigma_s$ is almost-sure winning if and only $\hat{\sigma}_s$ is almost-sure winning can be proved by Lemma~\ref{lemma:aswinning}. As a result, $\hat{G}$ satisfies Lemma~\ref{lemma:prop1}. As $\hat{\sigma}_s$ is almost-sure winning in $\hat{G}$, $\sigma_{s,\varepsilon}$ in Step~\ref{step:memoryless} is a memoryless solution to Problem~\ref{problem:suboptimal}.

\begin{remark}
Algorithm~\ref{alg:suboptimal} is not guaranteed to output a memoryless solution to Problem~\ref{problem:suboptimal} if one exists for the given $\varepsilon$. The output of Algorithm~\ref{alg:suboptimal} is memoryless only if the condition in Theorem~\ref{thm:arbitrarye} holds. If there \emph{exists an} $\varepsilon'>0$ such that no memoryless solutions exist, the condition in Theorem~\ref{thm:arbitrarye} is violated and Algorithm~\ref{alg:suboptimal} outputs a finite-memory solution to Problem~\ref{alg:suboptimal}, even if there exists a memoryless solution for the given $\varepsilon$.
\end{remark}

\section{Conclusion}

We considered the synthesis of optimal and $\varepsilon$-optimal almost-sure winning strategies in two-player turn-based stochastic games with Rabin winning conditions and discounted performance criteria. We showed that memoryless strategies suffice for being both optimal and almost-sure winning for the system and provided with an algorithm to solve one if they exist. We also showed a sufficient and necessary condition of the existence of memoryless $\varepsilon$-optimal almost-sure winning system strategies for all $\varepsilon > 0$ simultaneously. 
Given a specific $\varepsilon$, we proposed an algorithm which solves a memoryless $\varepsilon$-optimal almost-sure winning strategy if this condition is satisfied, and a finite-memory $\varepsilon$-optimal almost-sure winning strategy if this condition is violated.


%

\section*{Acknowledgements}
We appreciate R{\"u}diger Ehlers and Guillermo A. P{\'e}rez for the helpful discussions. 

\bibliographystyle{plain}
\bibliography{discountedRabin}

\end{document}